# Ordered and tunable Majorana-zero-mode lattice in naturally strained LiFeAs

Meng Li[1,2#], Geng Li[1,2,3,4#], Lu Cao[1,2#], Xingtai Zhou[1,2], Xiancheng Wang[1,2], Changqing Jin[1,2,4], Ching-Kai Chiu[5], Stephen J. Pennycook[1,2], Ziqiang Wang[6*], and Hong-Jun Gao[1,2,3,4*]

[1] Beijing National Center for Condensed Matter Physics and Institute of Physics, Chinese Academy of Sciences, Beijing 100190, PR China

[2] School of Physical Sciences, University of Chinese Academy of Sciences, Beijing 100190, PR China

[3] CAS Center for Excellence in Topological Quantum Computation, University of Chinese Academy of Sciences, Beijing 100190, PR China

[4] Songshan Lake Materials Laboratory, Dongguan, Guangdong 523808, PR China

[5] RIKEN Interdisciplinary Theoretical and Mathematical Sciences (iTHEMS), Wako, Saitama 351-0198, Japan

[6] Department of Physics, Boston College, Chestnut Hill, MA 02467, USA

[#]These authors contributed equally to this work.

[*]Correspondence to: hjgao@iphy.ac.cn; wangzi@bc.edu

**Majorana zero modes (MZMs) obey non-Abelian statistics and are considered building blocks for constructing topological qubits[1,2]. Iron-based superconductors with topological band structures have emerged as promising hosting materials, since isolated candidate MZMs in the quantum limit have been observed inside the topological vortex cores[3-9]. However, these materials suffer from issues related to alloying-induced disorder, uncontrolled vortex lattices[10-13] and a low yield of topological vortices[5-8]. Here, we report the formation of an ordered and tunable MZM lattice in naturally-strained stoichiometric LiFeAs by scanning tunneling microscopy/spectroscopy (STM/S). We observe biaxial charge density wave (CDW) stripes along the Fe-Fe and As-As directions in the strained regions. The vortices are pinned on the CDW stripes in the As-As direction and form an ordered lattice. We detect more than 90 percent of the vortices to be topological and possess the characteristics of isolated MZMs at the vortex center, forming an ordered MZM lattice with the density and the geometry tunable by an external magnetic field. Remarkably, with decreasing the spacing of neighboring vortices, the MZMs start to couple with each other. Our findings provide a new pathway towards tunable and ordered MZM lattices as a platform for future topological quantum computation.**

The long quest for Majorana excitations[14-19] has been fueled by recent experimental progress in material platforms where the localized MZM has been observed[16,20-23]. Among these platforms, iron-based superconductors are considered promising for the observation of clean and robust MZMs[3-5]. FeTe$_{0.55}$Se$_{0.45}$ was the first iron-based superconductor found to possess topological surface states (TSS)[24,25]. The large $\frac{\Delta^2}{E_F}$ ratio, where $\Delta$ is the superconducting (SC) energy gap and $E_F$ is the Fermi energy, allows the observation of an isolated MZM in the center of a topological vortex[4], which exhibits nearly quantized plateaus in differential conductance[9]. Later on, stoichiometric iron pnictides were investigated and the existence of MZMs has been verified in CaKFe$_4$As$_4$ (ref. [8]) and LiFeAs (ref. [6]). Very recently, 1D dispersing Majorana modes and MZMs have been observed in the layered 2D superconductors with topological band structures[26-28]. However, these platforms suffer from alloying-induced disorder, uncontrolled vortex lattice and a low yield of topological vortices, hindering their potential applications.

We start with LiFeAs crystal due to its homogeneous bulk electronic structure, well-defined cleaving surface and rich topological band structures[29]. LiFeAs belongs to the tetragonal crystal system, with lattice constants $a=b=3.8$ Å and $c=6.3$ Å (Fig. 1a). The As-As atomic direction defines the high-symmetry lattice orientations ([10] and [01] in Fig. 1b) of the mechanically cleaved nonpolar surface (Extended Data Fig. 1). The existence of strain in the LiFeAs lattice leads to the formation of wrinkles[30] or electronic ordering such as CDWs[31]. We identified such strained regions with a biaxial CDW phase (Fig. 1c), which appears as a coexistence of small periodic (2.7±0.2 nm) stripe patterns orienting along the Fe-Fe ([11]) direction ($CDW_{Fe-Fe}$) and large periodic (24.3±0.9 nm) stripe patterns orienting along the As-As ([10]) direction ($CDW_{As-As}$). To corroborate the effect of the local strain on the formation of the CDWs, we performed atomically-resolved STM imaging of the surface regions with and without the CDWs (Extended Data Fig. 2, Table S1). The distortion of the lattice in the CDW regions reveals the existence of biaxial and shear strain. The observed CDWs are reproducible in different LiFeAs samples and using different STM instruments (Extended Data Fig. 3). Both CDW wavevectors can be clearly resolved in the Fourier transformed (FT) image (Fig. 1d), with an angle of 45±1°, breaking the $C_4$ rotation and reflection symmetry of the crystal lattice. Interestingly, the bright stripes of both $CDW_{Fe-Fe}$ and $CDW_{As-As}$ show abnormal splitting and recombination behavior with increasing bias voltages near $E_F$ (Extended Data Fig. 4), which are accompanied by a π-phase shift in the stripe pattern (Extended Data Fig. 4c,d,k,p). The similar energy-dependent evolution of the stripes indicates that the two CDW phases are coupled with each other. We note that both CDW wavevectors are nondispersive with energy (Extended Data Figs. 4,5), precluding other possibilities such as quasi-particle interference[32,33].

To investigate the influence of the biaxial CDW phase on the SC behavior, we measure the d$I$/d$V$ spectra in different regions. In the unstrained region (black dot in Fig. 1b), the d$I$/d$V$ spectrum shows the multigap feature of LiFeAs, with a large gap of ~ 5.8 meV and a small gap of ~ 2.9 meV (black curve in Fig. 1e), consistent with previous reports[10,32,34,35]. In the strained region, the SC gap on the bright stripe (2.1±0.1 mV) of $CDW_{As-As}$ is smaller than that off the stripe (3.4±0.2 mV) (brown and yellow curves in Fig. 1e). Importantly, the two CDW orders affect the local superconductivity in very different ways. While the $CDW_{As-As}$ clearly modulates the SC gap size (Fig. 1f, left panel), the $CDW_{Fe-Fe}$ only modulates the intensity of the SC coherence peak without altering the gap size (Fig. 1f, right panel). The modulation of the SC gap suggests a strong coupling of the CDW to the superconductivity. We note that the periodicity of $CDW_{As-As}$ is significantly larger than the SC coherence length (4.8 nm, ref. [36]) in LiFeAs, and the modulation of SC behavior at such large scale is rare. To provide a straightforward visualization of the coupling between the CDW and the superconductivity, we conduct a series of d$I$/d$V$ maps at different energies and extract the SC gap values across the region in Fig. 1g. The SC gap map (Fig. 1h) shows clearly the modulation of the SC gap size by $CDW_{As-As}$.

We then apply an external field of 0.5 T normal to the sample surface, and find magnetic vortices emerging exclusively on the bright stripes of $CDW_{As-As}$ (Fig. 2a). The strong pinning effect of the vortices by the $CDW_{As-As}$ stripes is likely due to the suppressed SC gap in the bright As-As stripe regions. We take d$I$/d$V$ spectra across a typical vortex (along the red arrow in Fig. 2a), and find sharp zero-bias conductance peaks (ZBCPs) near the vortex center in the waterfall plot (Fig. 2b) and intensity map (Fig. 2c), together with a novel "X" like feature and discrete peaks on the two sides of the ZBCP. The ZBCP decays without splitting when the tip moves away from the vortex center (Fig. 2c), which is reminiscent of a MZM in the vortex core of other iron-based superconductors with TSS[4,6,8]. To better differentiate the energy positions of the discrete peaks, we show the negative 2$^{nd}$ derivative of the waterfall plot (Fig. 2d) and the intensity map (Fig. 2e) of the d$I$/d$V$ curves in Fig. 2b,c. A series of discrete vortex core states can be resolved (Fig. 2d,e), suggesting that the vortex is in the quantum limit[4,37]. The spatial evolution of the peak energy of the core states roughly follow straight lines (dashed drop lines in Fig. 2e), as expected for the low-lying Caroli-de-Gennes-Matricon (CdGM) bound states[38-40]. The other vortices exhibit similar spectroscopic features (Extended Data Fig. 6).

In order to determine the energies of the vortex core states, five individual d$I$/d$V$ spectra close to the center of the vortex core (red arrows in Fig. 2d) are replotted in Fig. 2f. The discrete vortex core states manifest as peaks marked by the vertical arrows. The CdGM states are bound states inside the SC gap and spatially localized in a vortex core. In an ordinary vortex, the energy of the bound states is half-integer quantized, $E_j = j \times E_0$, where j=±1/2, ±3/2… is the total angular momentum quantum number. However, for a topological vortex, the topological Dirac fermion surface states contribute an additional π phase, leading to a half-integer shift in the quantum number of the total angular momentum. As a result, the energy of the CdGM states is integer-quantized, $E_j = j \times E_0$, where j=0, ±1, ±2…, in the topological vortex. The average values and the deviations of the core states are summarized in Fig. 2g. Five discrete states at -1.76±0.11 ($L_{-2}$), -0.81±0.09 ($L_{-1}$), 0.00±0.06 ($L_0$), 0.86±0.10 ($L_1$), and 1.80±0.13 ($L_2$) meV are extracted, which approximately follow an integer-quantized sequence. Different from conventional CdGM states, the vortex core states exhibit weak spatial dispersions (horizontal black arrows in Fig. 2e and the extracted core states in Fig. 2g). To accurately determine the energy of the $L_0$ state, we decompose the original d$I$/d$V$ spectra into multiple Gaussian-like peaks[7,40] (Extended Data Fig. 7 and left panel in Fig. 2h). Utilizing this approach, we calibrate the $L_0$ states (right panel in Fig. 2h), and find that they locate at the Fermi energy. The robust ZBCP with a decaying intensity away from the vortex center is strong evidence of a MZM in a topological vortex[4]. The full width at half maximum (FWHM) of the fitted ZBCPs are ~ 0.4 meV, which is comparable to the energy resolution of our instrument (0.3 meV). These results strongly support the existence of MZMs in the vortex centers.

We then investigate the influence of the CDW stripes on the d$I$/d$V$ spectra of the vortices for further understanding of the spatial evolution of the CdGM states. In the negative 2$^{nd}$ derivative of the intensity maps of d$I$/d$V$ spectra across a topological vortex (Extended Data Fig. 8a), the dispersion happens at positions marked by the horizontal white dashed lines (left panel in Extended Data Fig. 8a). Remarkably, these horizontal lines are in good agreement with the positions of the dark Fe-Fe stripes (see height profile in the right panel in Extended Data Fig. 8a), indicating that the CDW$_{Fe-Fe}$ has strong influence on the vortex core states. The CdGM states are usually understood as eigenstates of a vortex under rotation symmetry. Once the rotation and reflection symmetries are broken, hybridization between the eigenstates can cause the spatial dispersions of the CdGM states (Fig. 2g and Extended Data Fig. 8b). We note that this is the first experimental observation of hybridization between vortex core states. The MZMs, however, do not disperse and remain at zero-energy, consistent with its topologically protected robustness against local perturbations.

Increasing the external field to 3 T, we observe that more vortices emerge, pinned to the bright CDW$_{As-As}$ stripes (Extended Data Fig. 9). Figure 3a shows the intensity map of d$I$/d$V$ spectra across a topological vortex in the upper panel and an ordinary vortex in the lower panel. Remarkably, its negative 2$^{nd}$ derivative shows a half-integer level shift in the energy of the vortex core states (Fig. 3b), which was considered as strong evidence distinguishing the topological vortex and the ordinary vortex[40]. We also plot the individual d$I$/d$V$ spectra at the two vortex cores as marked by the red and black dashed lines in Fig. 3a. A sharp ZBCP is seen in the topological vortex (upper panel in Fig. 3c). However, the ZBCP is absent for the ordinary vortex (lower panel in Fig. 3c). We measure the d$I$/d$V$ spectra at the center of each topological vortex as shown in Extended Data Fig. 10. All spectra show similar profiles, suggesting the electronic structure of the vortices are uniform. More than 90 percent of vortices are topological vortices, giving rise to an ordered MZM lattice (Extended Data Fig. 10). This conclusion holds for different magnetic fields from 0.5 to 6 T (see Methods, Extended Data Fig. 11 and Extended Data Fig. 12).

Now we discuss the possible origin of the MZM lattice. The topological nature of the pristine LiFeAs is very intriguing and contains two parts. First there is a set of helical Dirac fermion TSS akin to that in a 3D topological insulator, which are located closest to the Fermi level. Second and unique to LiFeAs, there is a set of bulk Dirac fermion states sitting ~ 10 meV above the Fermi level, which is characteristic of a topological Dirac semimetal. This dual topological nature complicates the vortex core spectrum for the unstrained LiFeAs, which appeared gapless, and prohibited the identification of the discrete vortex core states[6,10,41,42]. To uncover the electronic structure of the strained LiFeAs, we take wide energy scale d$I$/d$V$ spectra in the unstrained (lower region in Extended Data Fig. 13) and the CDW$_{Fe-Fe}$ (upper left region in Extended Data Fig. 13) regions, and in the biaxial CDW region on and off the As-As stripes (Fig. 3d and Extended Data Fig. 15). In the unstrained region, a hump can be recognized at an energy of ~ 33 meV (black curves in Fig. 3d and Extended Data Fig. 14), which is assigned to the top of $d_{xy}$ band[6]. In the strained regions, however, the hump feature disappears (brown, yellow and blue curves in Fig. 3d and Extended Data Fig. 14), indicating a renormalization in the local electronic structure by the CDW orders. In the uniaxial CDW$_{Fe-Fe}$ region, the coherence peak of the CDW$_{Fe-Fe}$ is identified at ~ 10 mV (blue arrow in Fig. 3d), in accordance with previous reports[31]. In the biaxial CDW region, additional peaks indicated by the brown and yellow arrows can be resolved (Fig. 3d and Extended Data Fig. 14). This new peak corresponds to the coherence peak of the CDW$_{As-As}$ order, since it is strongly modulated by the As-As stripes (Extended Data Fig. 15).

The above assigned CDW coherence peaks, together with the SC coherence peaks, lie on a background of a "V" shaped envelope in the d$I$/d$V$ spectra (Fig. 3e and Extended Data Fig. 16) with the dip locating at ~ -10 mV. This V-shaped envelope is a signature for the existence of a Dirac cone[4,43] in two dimensions. These observations together suggest that the most significant changes in the electronic structure in the strained biaxial CDW regions come from the breaking of the rotation and reflection symmetries, which gaps out the bulk Dirac cone protected by the rotation symmetry. This turns the local electronic structure of the stoichiometric, but strained biaxial CDW regions into a strong topological insulator exclusively, hosting the prominent helical Dirac fermion TSS (Fig. 3f) and generates the robust topological Majorana vortices in the SC state. This conclusion is corroborated by the fact that in the unstrained and the uniaxial CDW$_{Fe-Fe}$ regions, only ordinary vortices with no MZMs are observed (Extended Data Fig. 17).

We finally demonstrate the tunability of the ordered MZM lattice by external magnetic field. The ordered MZM lattice originates from the fact that all vortices are firmly pinned to the periodic bright stripes along the As-As direction (Fig. 4). At 0.5 T, the vortices have relatively low density and arrange themselves amorphously. Upon increasing the magnetic field to 5 T, the MZM lattice gradually evolves into the triangular lattice shape. It should be noted that at -0.5 mV, the vortices are ring-shaped, while at 0 and 0.5 mV they appear as solid circles. The d$I$/d$V$ maps reflect the wavefunctions of the bound states inside the vortex cores which are described by Bessel functions[38,44,45]. The ring structure for the negative energy branch suggests that the chemical potential lies above the Dirac point[8,45] (Fig. 3f). The pinning effect of the vortices by the stripes is so strong that the mobility of the vortices is limited to only one direction – along the stripe. By increasing the field, the newly emerged vortices are "doped" into the bright stripes, forming regular 1D vortex chains. The inter-chain vortex interactions are fixed due to the pinning effect, while the intra-chain vortex interactions can be manipulated by the applied magnetic field (Extended Data Fig. 18). The strong pinning effect provides better control over the vortex motion, as compared to the traditional two-dimensional disordered Abrikosov vortex lattices[10,11]. It is further demonstrated that the ordered lattice is tunable by the external magnetic fields. The overall size of the MZM lattice can be brought up to the micron level, only limited by the STM scanning capability, and could extend to an even larger area (bottom panel of Fig. 4). We have also obtained the correlation between the vortex spacing and the d$I$/d$V$ spectra of the vortices, indicating a coupling between the MZMs under high magnetic fields (Extended Data Fig. 19).

In summary, we have observed a micron-scale MZM lattice on naturally strained LiFeAs surface. The vortices are pinned to the $CDW_{As-As}$ stripes, giving rise to an ordered Abrikosov vortex lattice. We have demonstrated that the ordered lattice is tunable by external magnetic fields and observed the coupling between the MZMs under high magnetic fields. Our findings provide a promising platform for manipulating and braiding MZMs in the future. The large ordered array of MZMs can be suitable for realizing the "braiding by measurement-only" algorithm using interferometry[46].

# Methods

**Single-crystal growth**

High-quality LiFeAs single crystals were synthesized by using the self-flux method. A LiFeAs crystal was mounted on a STM sample holder in a glove box and transferred to an ultra-high vacuum chamber. The crystal was cleaved *in-situ*. The low cleaving temperature of 10 K is for a higher chance to find the strained regions hosting the CDWs compared with the room-temperature cleavage. After cleavage, the sample was immediately transferred to the STM scanner.

**STM/S experiments**

The STM/S measurements were conducted in an ultra-low temperature STM system equipped with 9-2-2 T vectorial magnets. Tungsten tips were etched chemically and calibrated on Au(111) surfaces before use. The d$I$/d$V$ spectra and maps were obtained by a standard lock-in technique with a modulation voltage of 0.1 mV at 973.0 Hz. All the images, spectra and d$I$/d$V$ maps were taken at 400 mK.

**Fitting method**

Linear least square fitting was used to determine the positions of the vortex core states. Five Gaussian peaks representing the in-gap bound states were used for the fitting of each spectrum. In the fitting process, all the parameters of the Gaussian peaks are set free. The peak positions of the bound states were extracted from the fitting results.

**Statistics of the topological vortices**

The determination of percentage of topological vortices in the biaxial CDW region was performed by the measurement of d$I$/d$V$ spectra of 51 vortices under 3 T. A relatively smaller region that contains 3 to 4 vortices within the area were mapped at zero energy so that the centers of the vortex cores could be accurately located. The spectra were all taken under the same scanning settings. The spectra were further calibrated by the multi-Gaussian peak fitting method to accurately locate the positions of the bound states. The same strategy was used for different fields from 0.5 to 6 T, and the topological vortices were always higher than 90% of the total number.

**Acknowledgements:** We thank G. Su and H. Ding for helpful discussions. The work is supported by the Ministry of Science and Technology of China (2019YFA0308500, 2018YFA0305700, 2017YFA0206303), the National Natural Science Foundation of China (61888102, 51991340, 52072401), the Chinese Academy of Sciences (XDB28000000, XDB30000000, 112111KYSB20160061), and the CAS Project for Young Scientists in Basic Research (YSBR-003). Z.W. is supported by the US DOE, Basic Energy Sciences Grant No. DE-FG02-99ER45747.

**Author Contributions:** H.-J.G. designed the experiments and supervised the project. X.W. and C.J. prepared samples. M.L., G.L., L.C. and X.Z. performed STM experiments with guidance of H.-J.G. G.L., C.-K.C, S.J.P., Z.W. and H.-J.G. did data analysis and wrote the manuscript. All of the authors participated in analysing experimental data, plotting figures, and writing the manuscript.

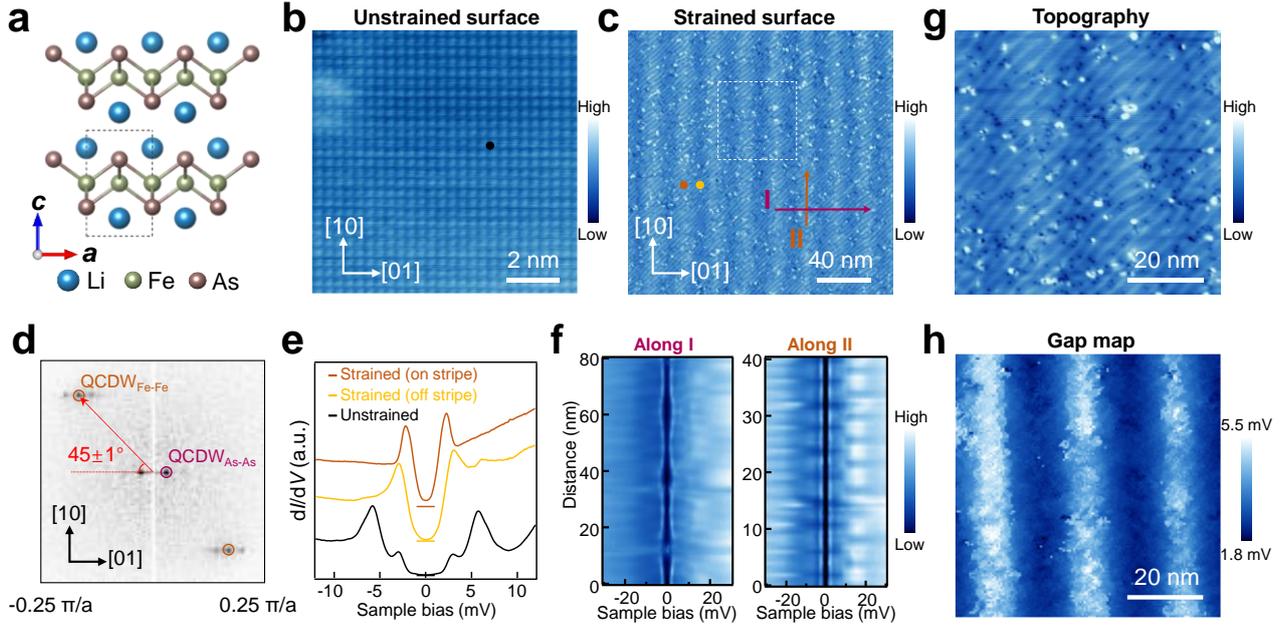

**Fig. 1 | Crystalline structure, topographic image and superconducting behavior of the unstrained and strained regions. a,** Atomic model of the LiFeAs single crystal. **b,c,** STM images of unstrained (**b**) and strained (**c**) regions on a cleaved LiFeAs single crystal, respectively (scanning settings for (**b**): bias $V_s$=-3 mV, setpoint $I_t$=400 pA; for (**c**): $V_s$=-30 mV, $I_t$=50 pA). **d,** FT image of **c**, showing the wavevectors of the $CDW_{Fe-Fe}$ and $CDW_{As-As}$. The angle between the wavevectors of $CDW_{Fe-Fe}$ and $CDW_{As-As}$ is 45±1°. **e,** Typical d$I$/d$V$ spectra taken at the bright stripe (brown curve), off the bright stripe (yellow curve) of the $CDW_{As-As}$ and in the unstrained region (black curve). The d$I$/d$V$ spectra are vertically offset for clarity, and the horizontal solid lines represent positions of zero density of states for each of the curves. The bright stripe region has smaller SC gap size and higher in-gap density of states compared with that of the off-stripe region. **f,** Intensity maps of the d$I$/d$V$ spectra along arrows I (left panel) and II (right panel) in **c**. The SC gap value is modulated by the large stripes of the $CDW_{As-As}$, while the $CDW_{Fe-Fe}$ pattern modulates only the height of SC coherence peaks. **g,** Zoom-in STM image of the white dashed box in **c** ($V_s$=-30 mV, $I_t$=50 pA). **h,** SC gap map of the same region in **g**, showing reduced SC gap sizes on the bright As-As stripes.

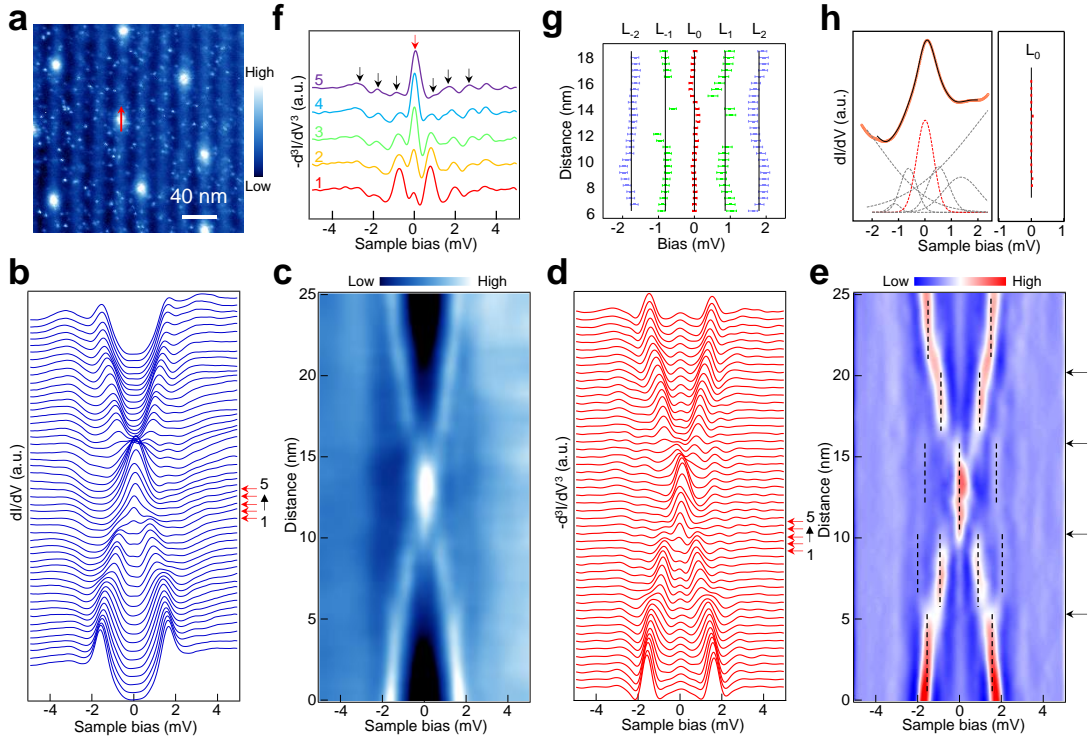

**Fig. 2 | d$I$/d$V$ map of the vortices and the analysis of the vortex bound states under 0.5 T. a**, d$I$/d$V$ map of the biaxial CDW region ($V_s$=0 mV), showing that the vortices are pinned to the bright As-As stripes. **b,c,** Waterfall plot (**b**) and intensity map (**c**) of the d$I$/d$V$ spectra along the red arrow in **a**, showing sharp ZBCP at the center of the vortex. **d,e,** Negative 2$^{nd}$ derivative of **b** and **c**, respectively, indicating that a series of discrete peaks locate on the two sides of the ZBCP. The dashed drop lines in **e** outline the positions of the discrete peaks. Following the dashed drop lines, the spatial evolvement of the discrete peaks is divided into sections, as marked by the horizontal black arrows. The peak positions show slight spatial dispersion from straight lines. **f,** Negative 2$^{nd}$ derivative of the 5 typical d$I$/d$V$ spectra (arrowed in red in **d**) taken close to the center of the vortex in **a**. The ZBCP is highlighted by a vertical red arrow, and the other discrete energy bound states are highlighted by vertical black arrows. **g,** Statistical analysis of the peak positions in **e**. Five energy bound states at -1.76±0.11 ($L_{-2}$), -0.81±0.09 ($L_{-1}$), 0.00±0.06 ($L_0$), 0.86±0.10 ($L_1$), and 1.80±0.13 ($L_2$) meV are extracted. **h,** Left panel: the black circles show a typical d$I$/d$V$ spectrum near the vortex core, with the extracted $L_0$ state from **d** deviating from zero energy. The red peak is the calibrated ZBCP by multi-Gaussian peak fitting. The grey lines indicate the fitted CdGM states and the background. The black solid line is the fitted curve. Right panel: spatial distribution of the calibrated $L_0$ state, showing robust ZBCPs.

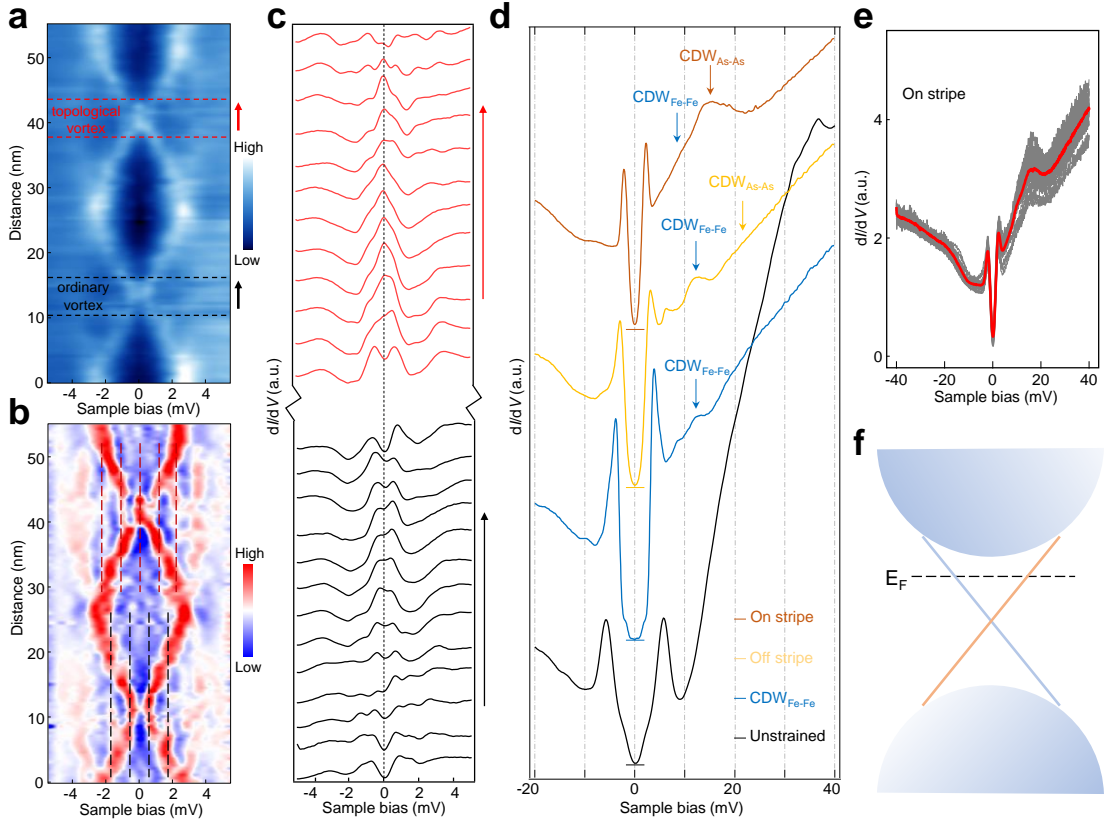

**Fig. 3 | Analysis of the d$I$/d$V$ spectra along a linecut across two neighboring vortices and the origin of the MZMs. a,** Intensity plot of d$I$/d$V$ spectra across two neighboring vortices along a CDW$_{As-As}$ stripe under 3 T. **b,** Negative 2$^{nd}$ derivative of **a**. The vertical dashed lines highlight the vortex core states. The upper vortex is topological with a ZBCP, while the lower vortex is ordinary. The discrete peaks of the ordinary vortex show half-integer level shift with that of the topological vortex. **c,** Zoom-in of the individual d$I$/d$V$ spectrum taken at regions close to the two vortex cores, as marked by the horizontal red and black dashed lines in **a**. The vertical black dashed line labels the zero bias. **d,** Wide-range d$I$/d$V$ spectra at the on As-As stripe (brown) and off As-As stripe (yellow) locations in a biaxial CDW region, the uniaxial CDW$_{Fe-Fe}$ region (blue), and the unstrained region (black), respectively. The horizontal solid lines highlight the positions of zero density of states for each curve. The big hump at ~ 33 mV of the black curve assigned to the band top of $d_{xy}$ in the unstrained region is absent in the uniaxial CDW$_{Fe-Fe}$ and biaxial CDW regions, suggesting significant band renormalization effects that also reduce the slope of the density states increase with energy. The coherence peaks of CDW$_{As-As}$ and CDW$_{Fe-Fe}$ are highlighted by vertical arrows. **e,** Averaged d$I$/d$V$ spectra taken on the bright As-As stripes in the biaxial CDW region. The linear dispersion of the TSS can be related to an approximately V-shaped spectral shape with the Dirac point located ~ -10 meV. **f,** Schematics of the band structure of the strained LiFeAs around the Γ point. The bulk states are gapped out by the biaxial CDW order, giving rise to the TI surface states.

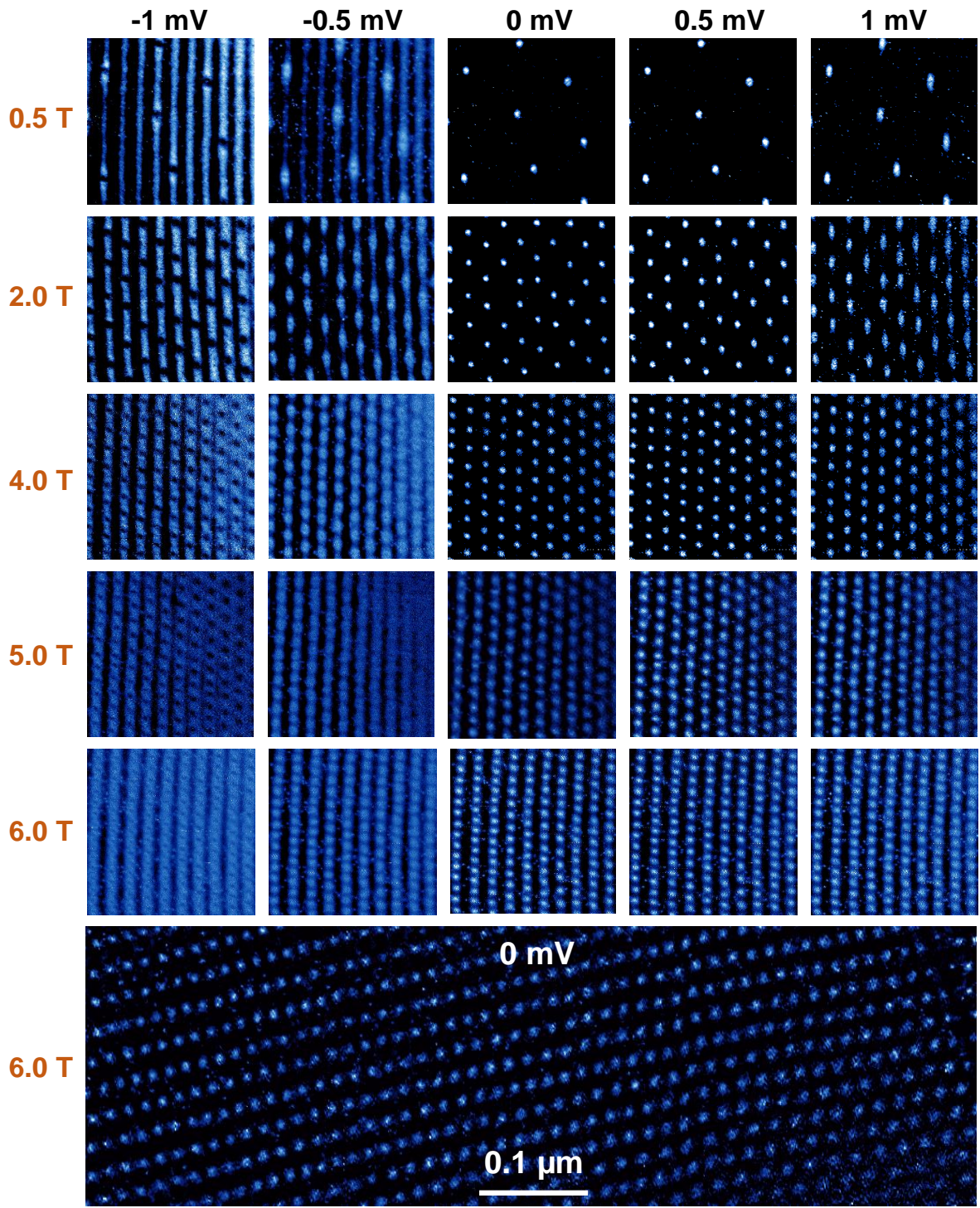

**Fig. 4 | Tuning the MZM lattice with external magnetic fields.** Series of d$I$/d$V$ maps of the MZM vortices in the biaxial CDW region under magnetic fields of 0.5 T, 2 T, 4 T, 5 T and 6 T and at bias voltages from -1.0 mV to 1.0 mV, respectively. Large-scale, ordered MZM lattice of different spacing and density is formed when increasing external magnetic fields. The scanning areas are 0.24 μm × 0.24 μm. Bottom panel: Micron-size ordered MZM lattice under 6 T. The scanning area is 0.26 μm × 0.82 μm.

# Unstrained region

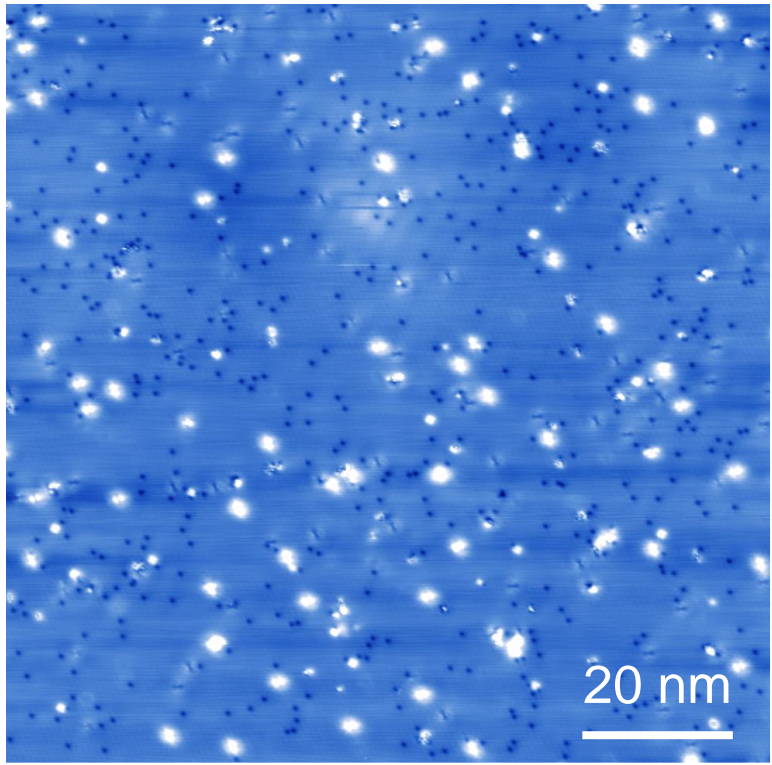

**Extended Data Fig. 1 | Large-scale STM image of the unstrained region on LiFeAs ($V_s$=-100 mV, $I_t$=50 pA).**

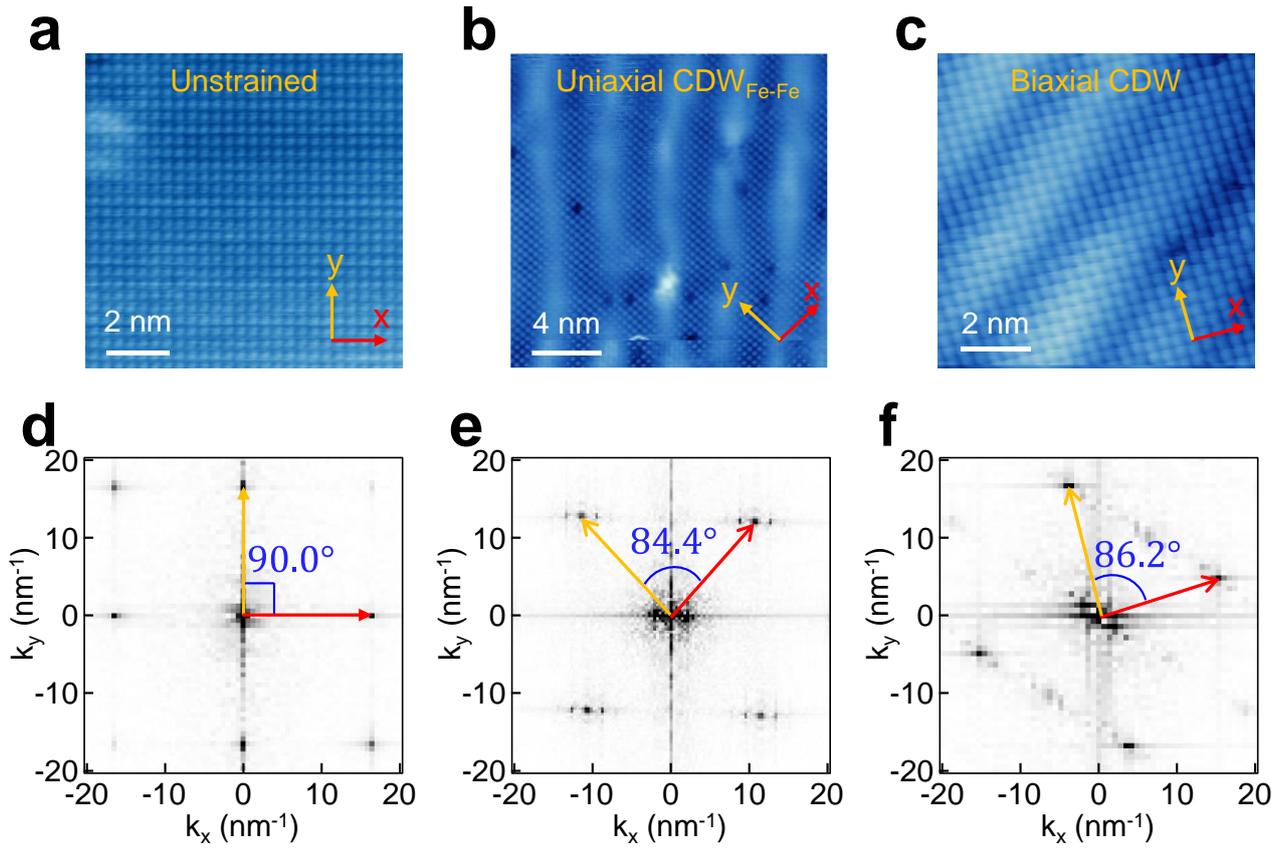

**Extended Data Fig. 2 | Comparison of lattice distortion among different regions of LiFeAs. a-c,** Atomic-resolution topographic images of the unstrained region, the uniaxial CDW$_{Fe-Fe}$ region and the biaxial CDW region. The crystallographic directions are shown in the lower right part. **d-f,** FT images of **a-c**. The angles of the crystallographic directions are marked by colored arrows.

| Region | x (Å) | y (Å) | $\Delta_a$ (%) | $\Delta_b$ (%) | θ (°) |
|---|---|---|---|---|---|
| Unstrained | 3.85 | 3.85 | 0.00 | 0.00 | 90.0 |
| Uniaxial CDW$_{Fe-Fe}$ | 3.96 | 3.71 | +2.86 | -3.64 | 84.4 |
| Biaxial | 3.95 | 3.67 | +2.60 | -4.68 | 86.2 |

**Table S1 | Lattice constants in the unstrained and the strained regions**

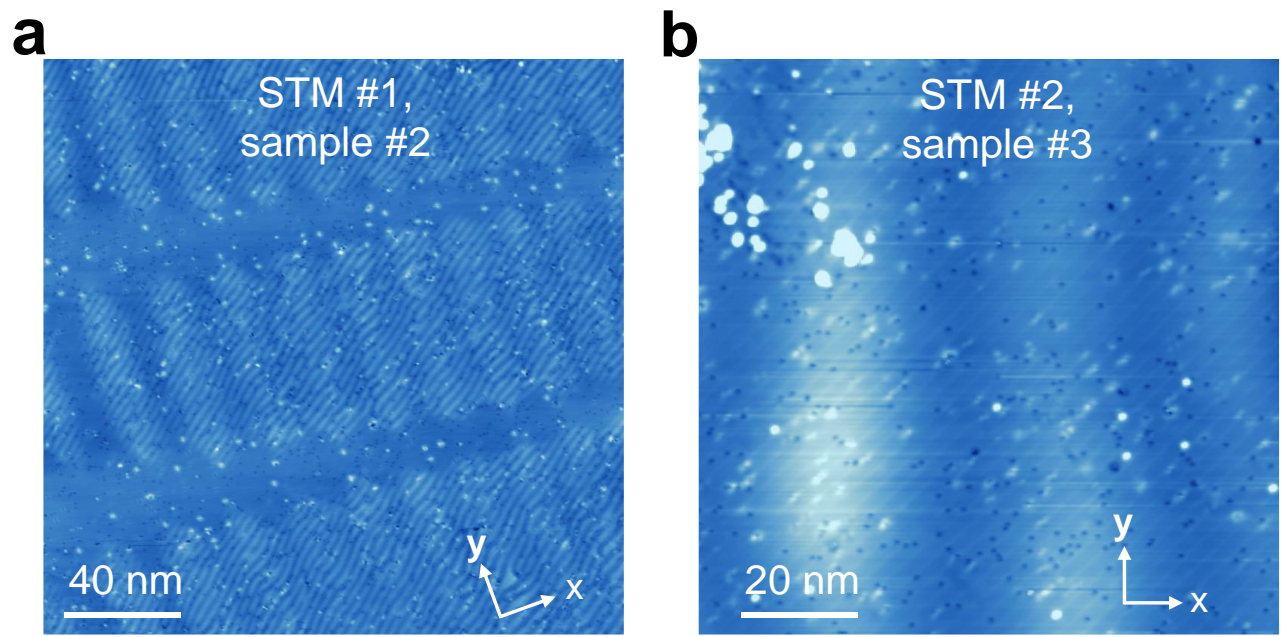

**Extended Data Fig. 3 | Reproducibility of the biaxial CDW results. a,** Topographic image of the biaxial CDW region, obtained on STM #1, sample #2. **b,** Topographic image of another biaxial CDW region, obtained on STM #2, sample #3.

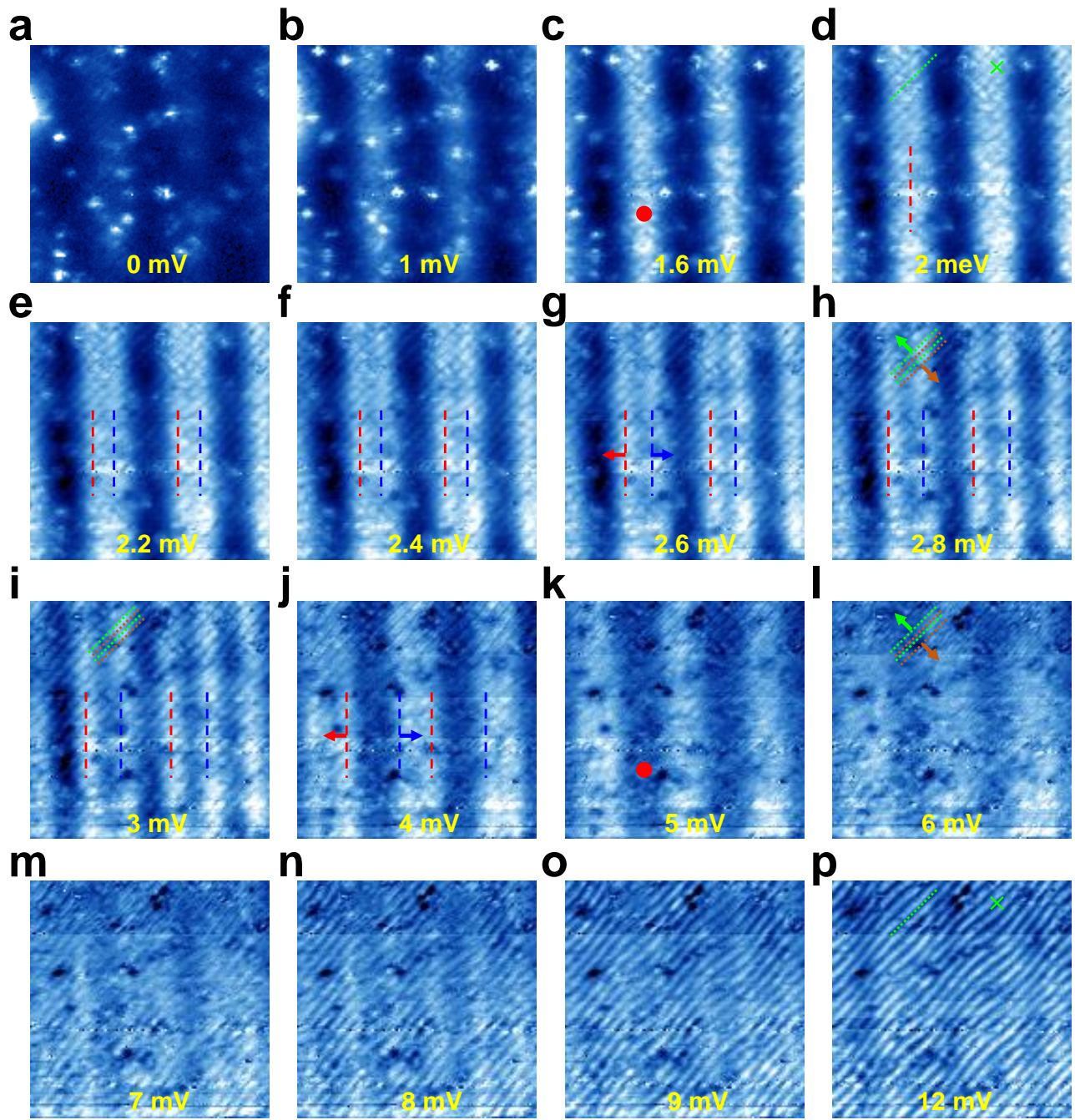

**Extended Data Fig. 4 | Evolution of the CDW$_{\text{As-As}}$ and CDW$_{\text{Fe-Fe}}$ stripes with bias voltages. a-p,** d$I$/d$V$ maps of a 70 nm × 70 nm biaxial CDW region under bias voltages of 0, 1.0, 1.6, 2.0, 2.2, 2.4, 2.6, 2.8, 3.0, 4.0, 5.0, 6.0, 7.0, 8.0, 9.0 and 12.0 mV, respectively. The red dot markers in c and k outline the $\pi$ phase shift of the CDW$_{\text{As-As}}$ pattern below and above the superconducting gap. The green cross markers in d and p outline the $\pi$ phase shift of the CDW$_{\text{Fe-Fe}}$ pattern below and above the superconducting gap. The CDW$_{\text{As-As}}$ pattern shows a splitting behavior of the stripes which starts at an

energy of ~2.2 mV (e). The stripes split into two sets as highlighted by red and blue dashed lines (e-j). Each set of the stripes keeps the same periodicity with the original $CDW_{As-As}$ stripes. However, both sets show dynamical behavior with energy by moving in opposite directions, as highlighted by the red and blue arrows (g,j). At an energy of ~4 mV, the split stripes recombine, returning back into a single set of stripes (j,k), with a $\pi$ phase shift (c,k). Similar splitting-recombining behavior exists in the $CDW_{Fe-Fe}$ stripes. The splitting starts at an energy of ~2.8 mV as highlighted by green and brown dotted lines (h). The recombination happens at ~ 9 mV which is accompanied with a $\pi$ phase shift (d,p). The same behavior happens on the negative bias side for the $CDW_{As-As}$ and $CDW_{Fe-Fe}$ stripes. The full set of d$I$/d$V$ maps can be found in Supplementary Video 1. The evolution of the uniaxial $CDW_{Fe-Fe}$ stripes can be found in Supplementary Video 2.

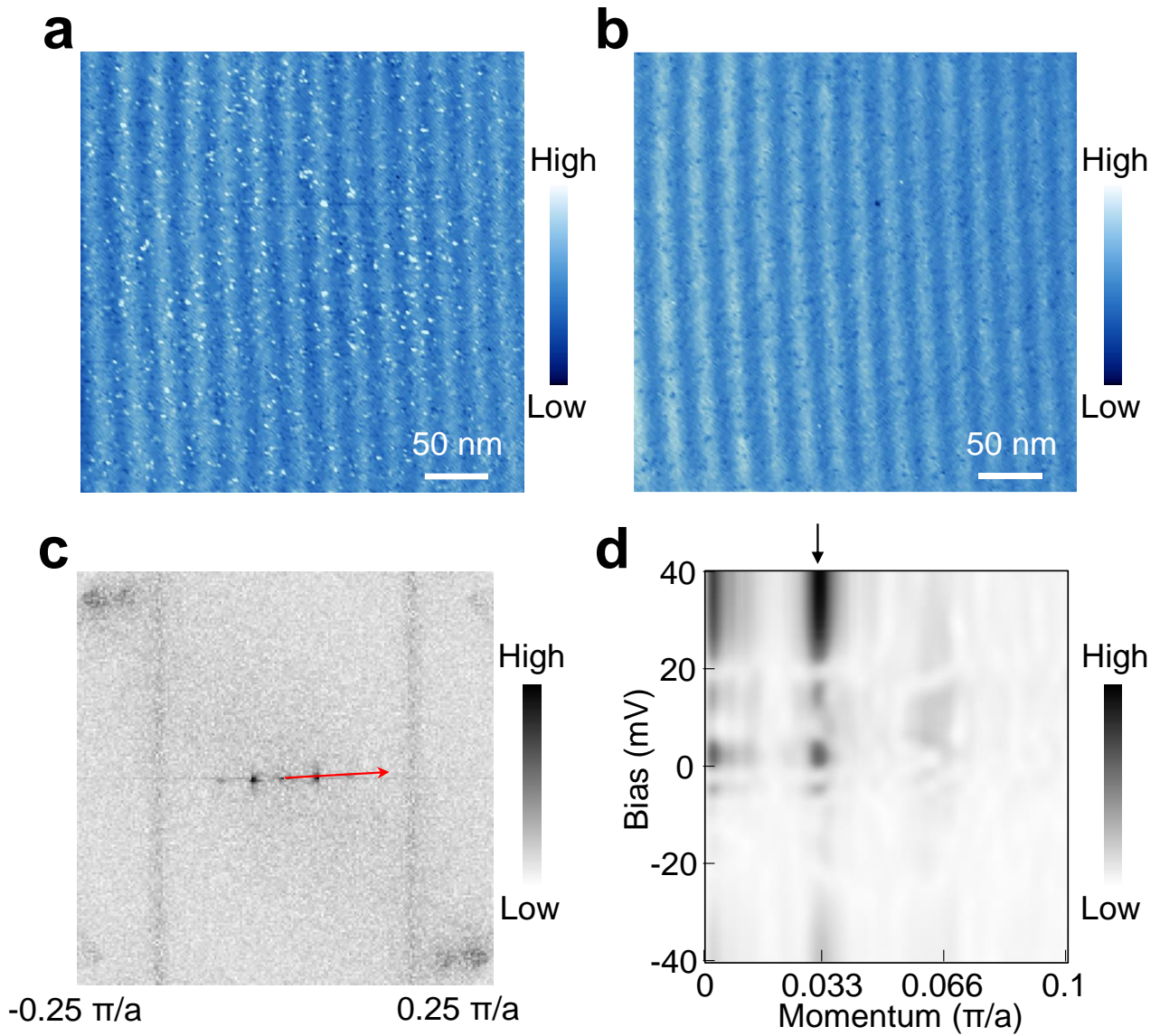

**Extended Data Fig. 5 | Static charge ordering of the As-As direction stripes under different bias voltages. a,b,** Topography (a) and d$I$/d$V$ map at 22 mV (b) of a 350 nm × 350 nm biaxial CDW region. **c,** FT image of b. **d**, Intensity plot of the FT images of d$I$/d$V$ maps of a under different energies along the red arrow in c. The black arrow highlights the wavevector of CDW$_{As-As}$. The periodicity of the stripes does not change under different bias voltages, indicating the CDW characteristic of the As-As stripes.

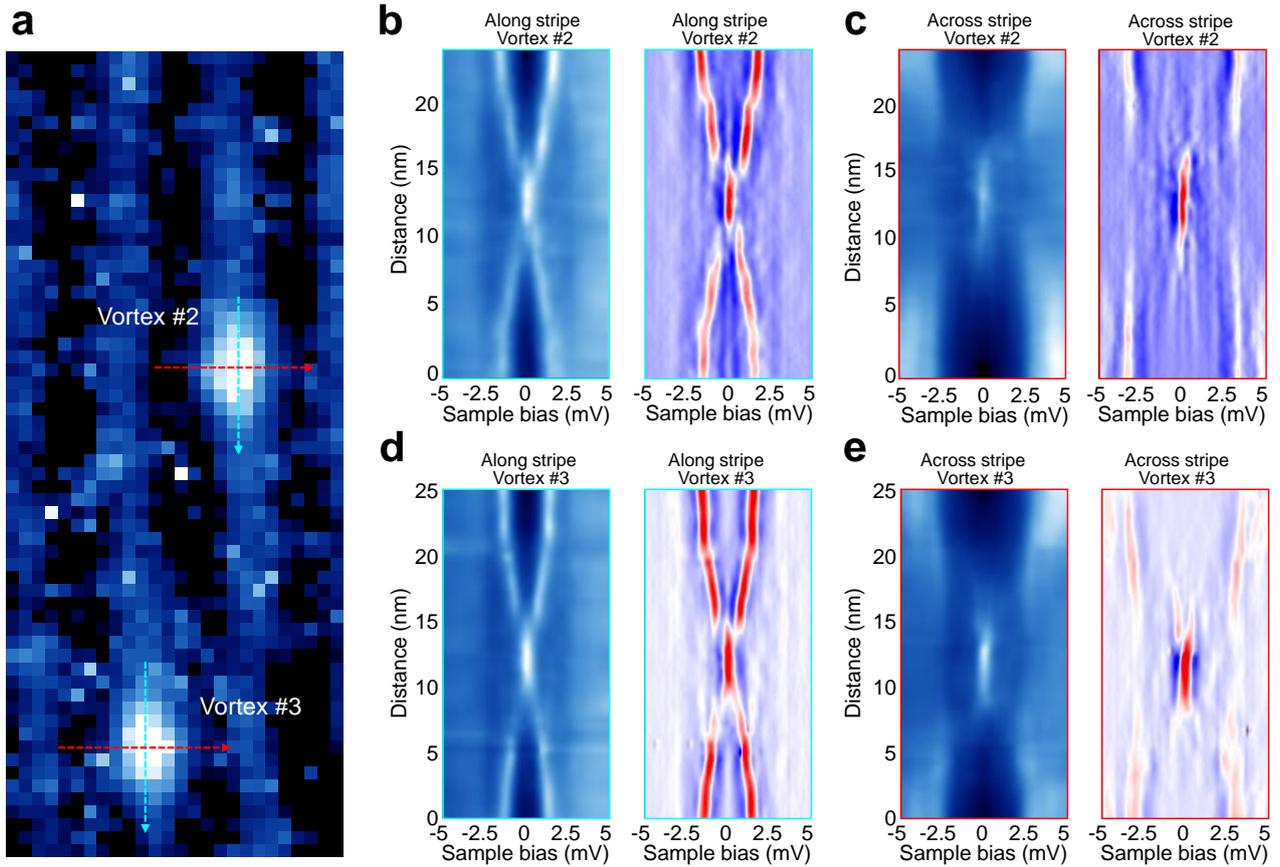

**Extended Data Fig. 6 | d$I$/d$V$ map of two vortices, d$I$/d$V$ spectra and the negative 2$^{nd}$ derivative of the linecuts along two perpendicular directions across the two vortices. a,** d$I$/d$V$ map of two vortices under 0.5 T. **b,d,** Intensity maps of d$I$/d$V$ spectra along (blue dashed arrows in a) the CDW$_{As-As}$ stripes (left panels) and the corresponding negative 2$^{nd}$ derivative of the maps (right panels) for vortex #2 (b) and vortex #3 (d). **c,e,** Intensity maps of d$I$/d$V$ spectra across (red dashed arrows in a) the CDW$_{As-As}$ stripes (left panels) and the corresponding negative 2$^{nd}$ derivative of the maps (right panels) for vortex #2 (c) and vortex #3 (e).

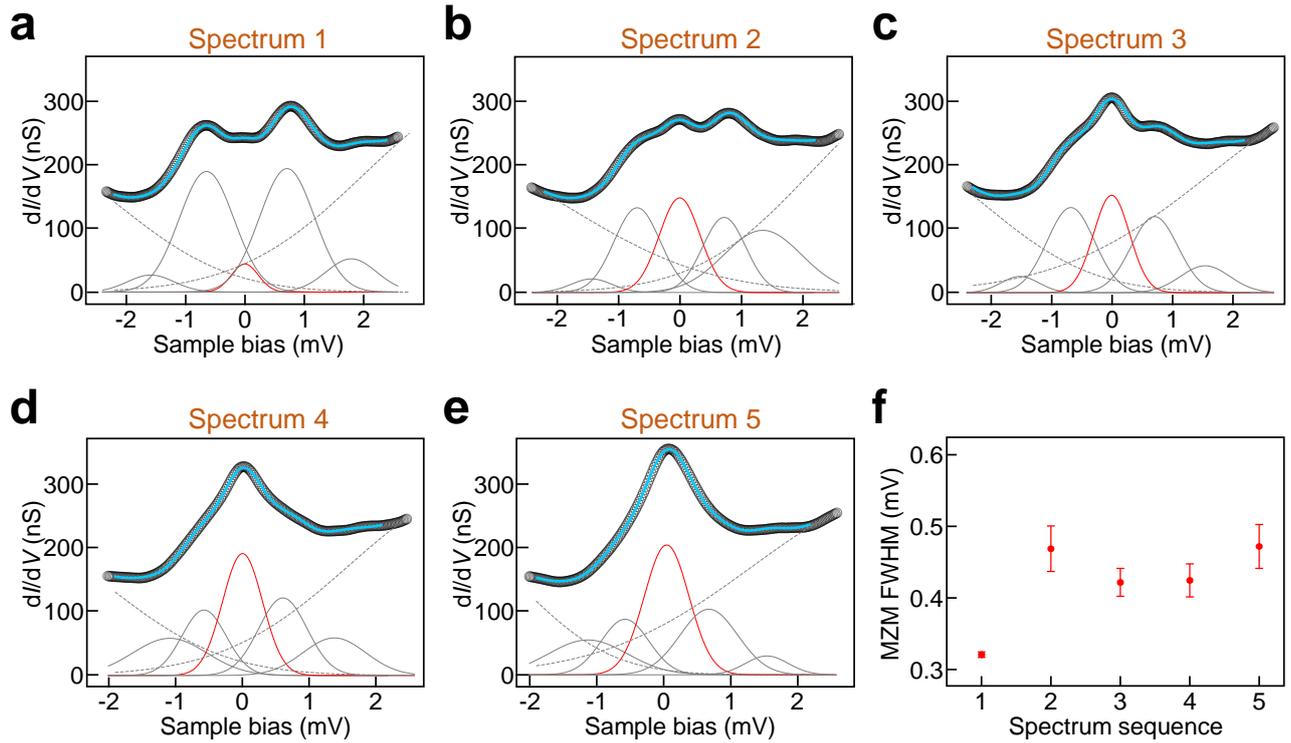

**Extended Data Fig. 7| Fitting of the MZM and low-order CdGM states of the d$I$/d$V$ spectra. a-e,** Fitting of the MZM and the 1$^{st}$ and 2$^{nd}$ order CdGM states of the five spectra indicated by arrows in Fig. 2b. The black circles are the experimental values after smoothing out the high-frequency noise. Five Gaussian peaks are used for the peak fitting. The fitted MZMs are shown by red lines, and the fitted CdGM states by gray lines. The gray dashed lines represent the background. Sharp ZBCPs are always seen in each of the spectra, suggesting that the topological vortex is in the quantum limit. The black curves are the fitted curves by adding the five fitted peaks and the background. **f,** Full width at half maximum (FWHM) of the fitted MZM peaks. The FWHM of the five MZMs are around 0.4 meV, in accordance with the energy resolution of our instrument (0.3 meV).

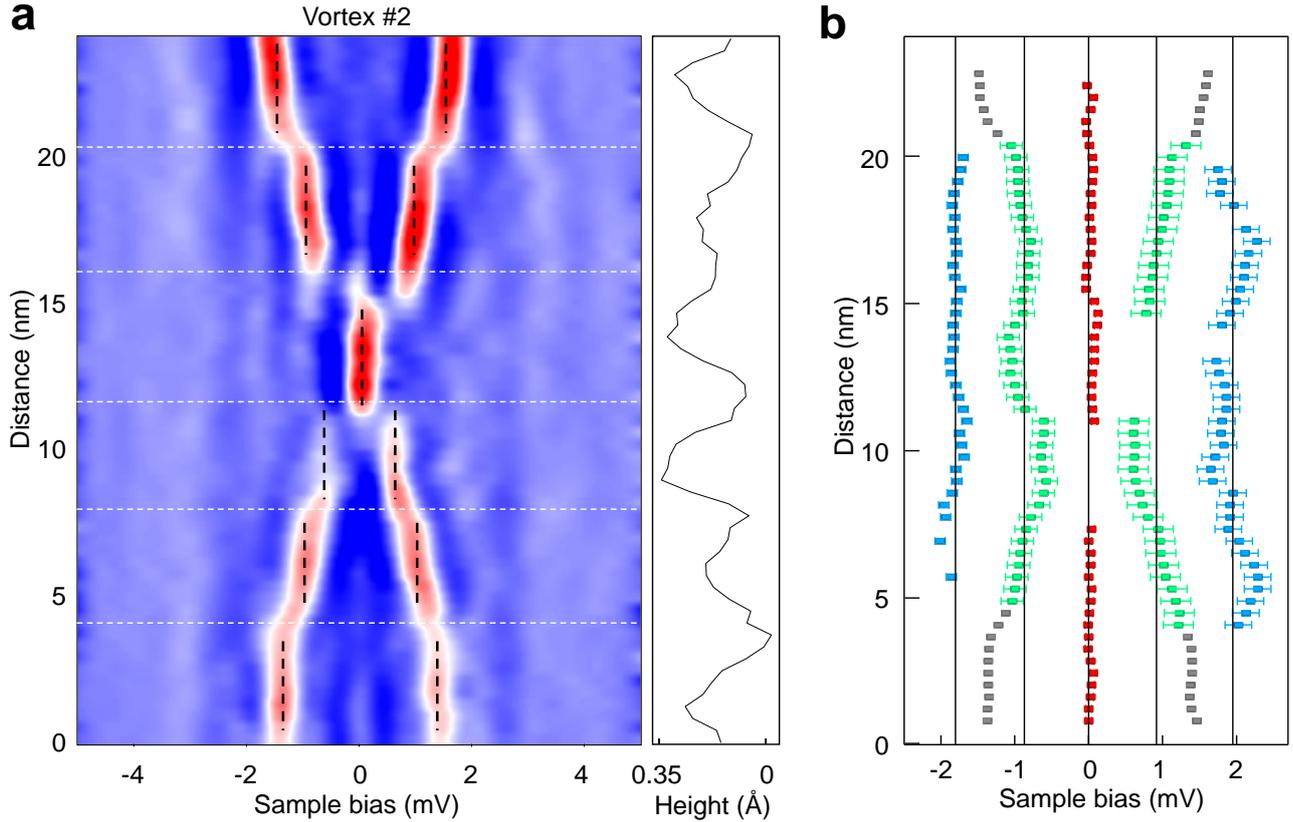

**Extended Data Fig. 8 | Detailed analysis of the negative 2$^{nd}$ derivative of the intensity plot of the d$I$/d$V$ spectra across vortex #2. a,** Left panel: Intensity map of -d$^3I$/d$V^3$ spectra across a topological vortex (vortex #2) along the As-As direction. The vertical black dashed lines outline the positions of the discrete vortex core states. The bound states show spatial dispersion of peak positions, similar to that observed in vortex #1. The horizontal white dashed lines outline the positions of dark stripes of CDW$_{Fe-Fe}$, in accordance with the height profile in the right panel. The spatial variations of the vortex bound states happen in vicinity to the positions of the dark stripes of CDW$_{Fe-Fe}$, suggesting a signature of the hybridization of the core states, which is a consequence of C$_4$ rotation and reflection symmetry breaking. **b,** Statistical analysis of the peak positions in a. Five energy bound states at -1.81±0.08 (L$_{-2}$), -0.87±0.15 (L$_{-1}$), 0.03±0.04 (L$_0$), 0.92±0.20 (L$_1$) and 1.95±0.19 (L$_2$) meV are extracted.

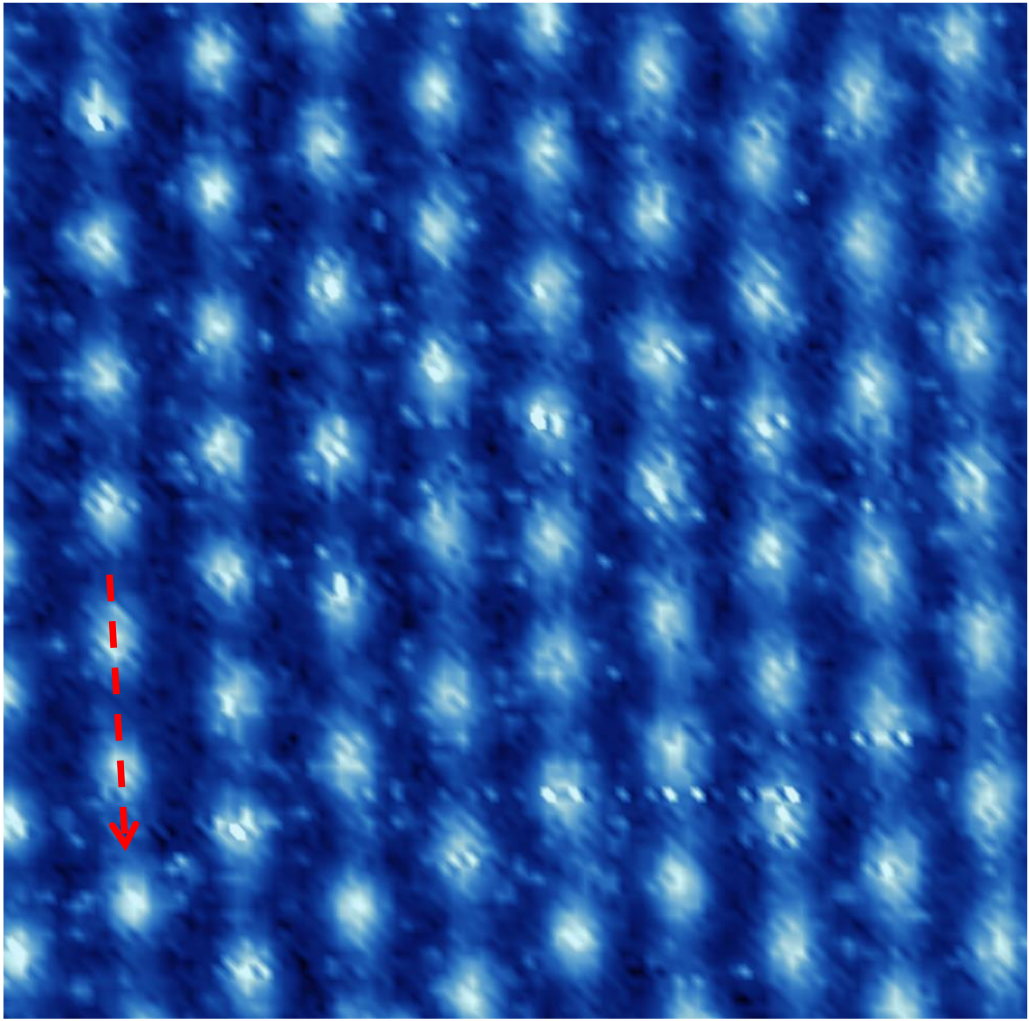

**Extended Data Fig. 9 | d$I$/d$V$ map of the biaxial CDW region under 3 T.** The map is taken at a bias voltage of 0 mV. The red dashed arrow outlines the path along which the d$I$/d$V$ spectra in Fig. 3a-c are taken.

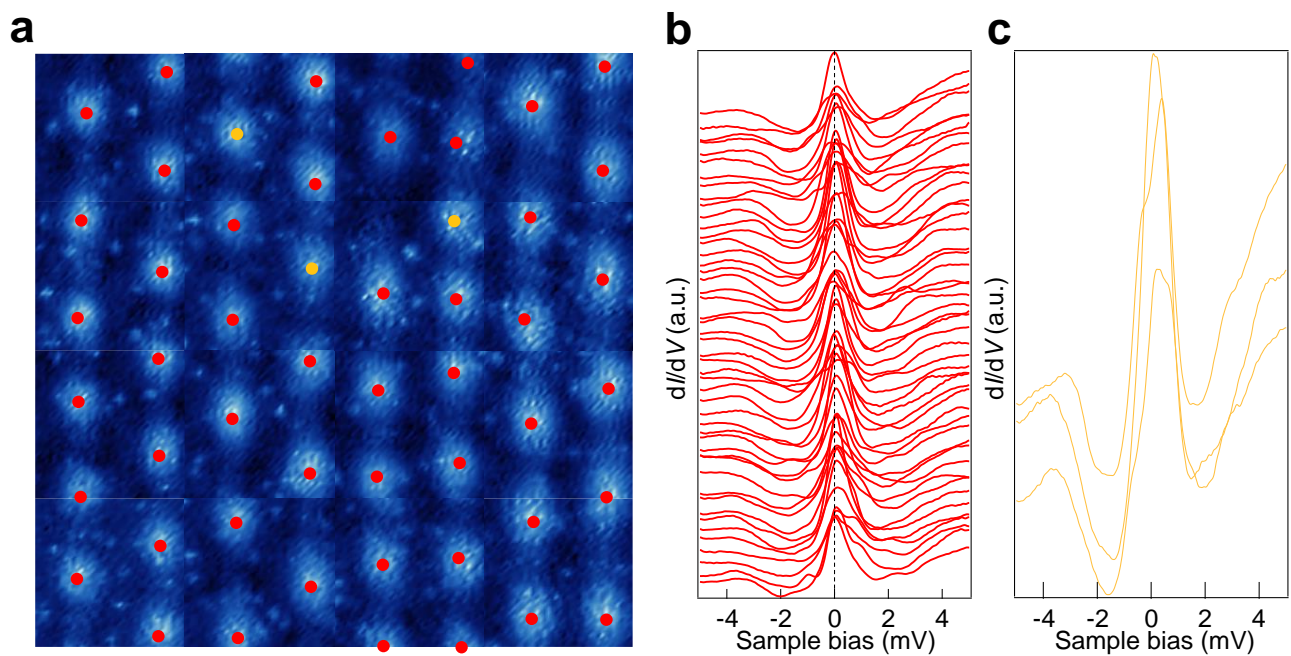

**Extended Data Fig. 10 | Statistics of the topological and ordinary vortices in the MZM lattice. a,** d$I$/d$V$ map of the large-scale vortices at 0 mV. The red dots mark the vortices with sharp ZBCP (topological) of the MZMs, and the yellow dots mark the vortices without sharp ZBCP (ordinary). The spectra are calibrated by the multi-Gaussian peak fitting to extract the accurate energy positions of the vortex bound states. 48 out of 51 vortices show the sharp ZBCP at the centers. The scanning area is 200 nm × 200 nm. **b,c,** Individual d$I$/d$V$ spectrum at each of the topological (b) and ordinary (c) vortices. Over 90 percent of the vortices have the characteristics of the MZM.

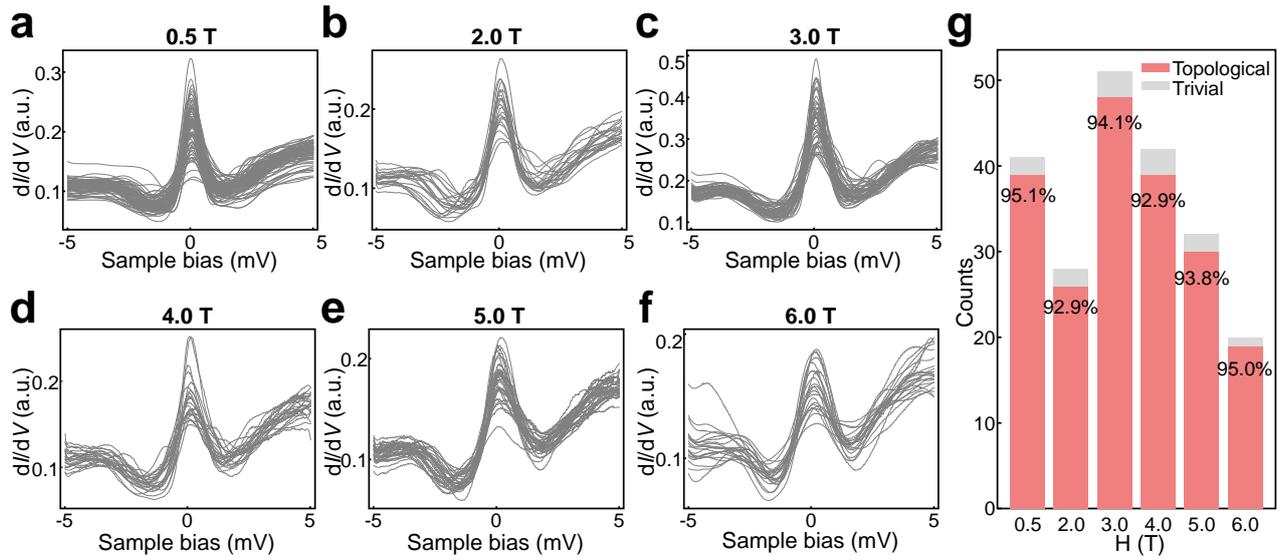

**Extended Data Fig. 11 | Statistics of Majorana zero modes under different magnetic fields. a-f,** d$I$/d$V$ spectra taken at the centers of different vortex cores under different magnetic fields. **g,** Histogram and percentage of topological vortices under different magnetic fields. The percentage of topological vortices is above 90% at all the magnetic fields up to 6 T.

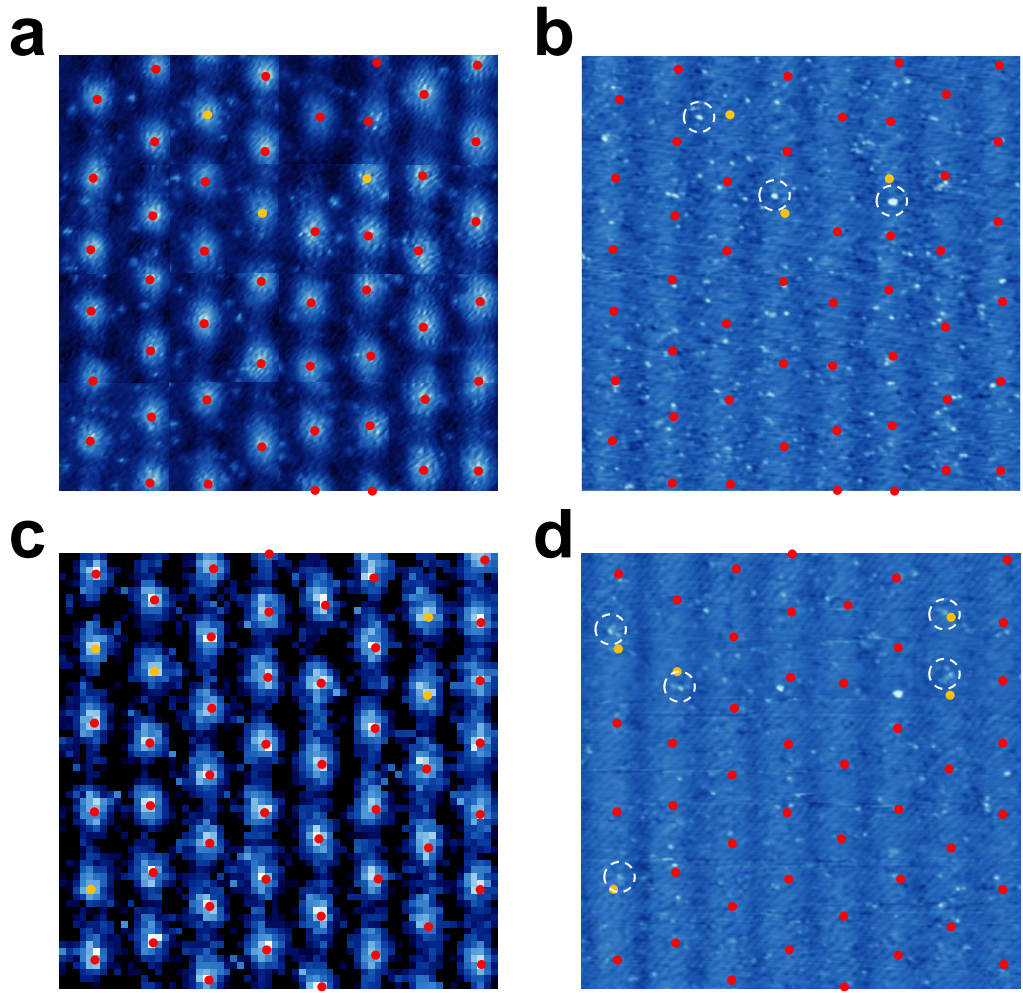

**Extended Data Fig. 12 | Possible defect-induced topological-trivial vortex transition. a,** Vortex lattice in the first round of measurement under a magnetic field of 3T. The red and yellow dots represent topological and trivial vortices, respectively. **b,** Topographic image of the same region in a. The positions of the topological and trivial vortices are overlaid. **c,** Vortex lattice of the second round of measurement after the field is ramped down to 0 and then back to 3 T. The red and yellow dots represent topological and trivial vortices, respectively. **d,** Topographic image of the same region in c. The positions of the topological and trivial vortices are overlaid. The white dashed circles in b and d mark the positions of the impurities. Although the trivial vortices appear in different regions for the two rounds of measurements, they are all located in the vicinity of the impurities ("brighter dots"), as outlined by the white dashed circles in b and d.

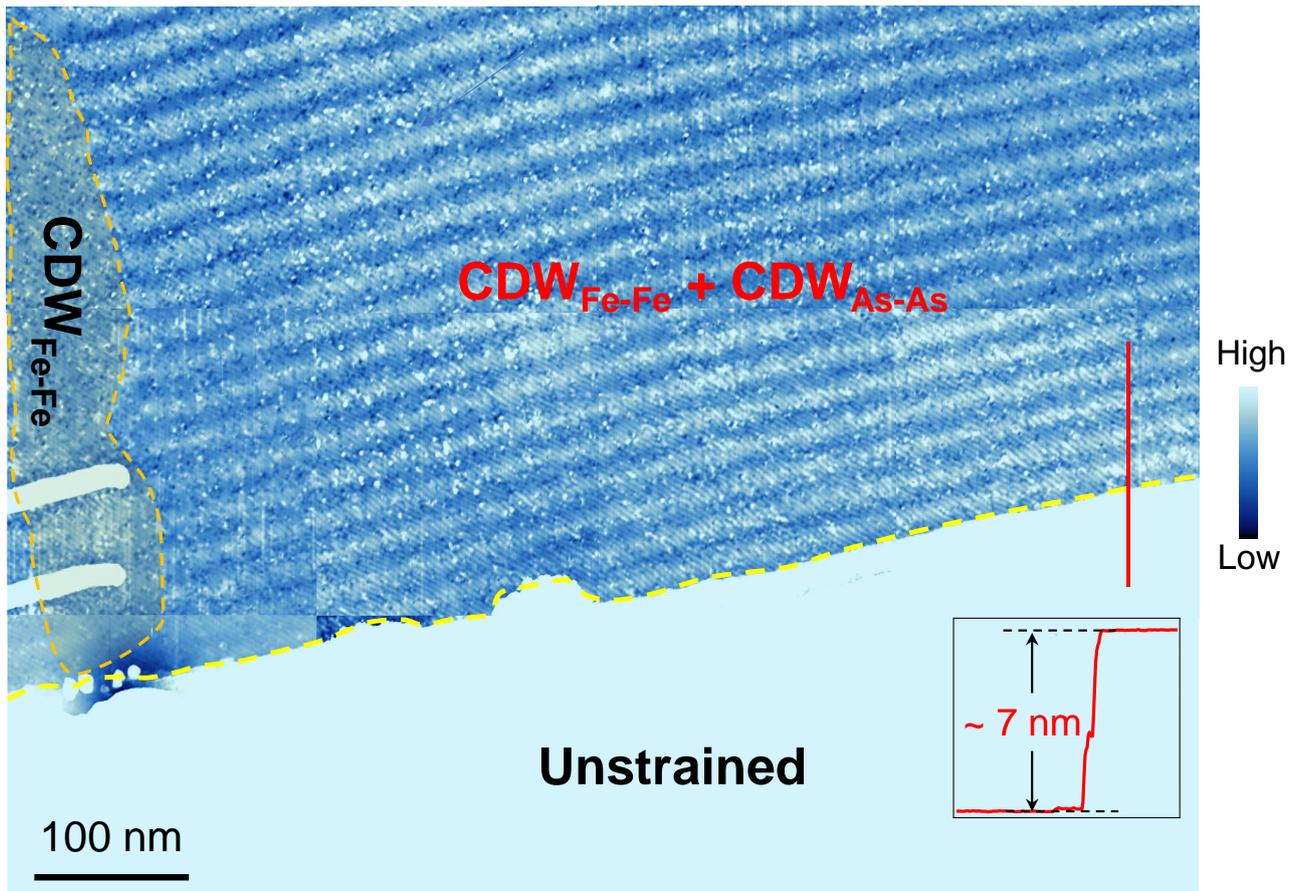

**Extended Data Fig. 13 | Large-scale STM image of the strained and unstrained regions of the LiFeAs.** The large-scale image is stitched from 12 independent STM topographic images. The strained region locates between two big steps with heights of ~ 7 nm, consisting of two kinds of regions, the uniaxial $CDW_{Fe-Fe}$ (upper left) and the biaxial CDW (upper right) regions, $V_s$=-20 mV, $I_t$=30 pA.

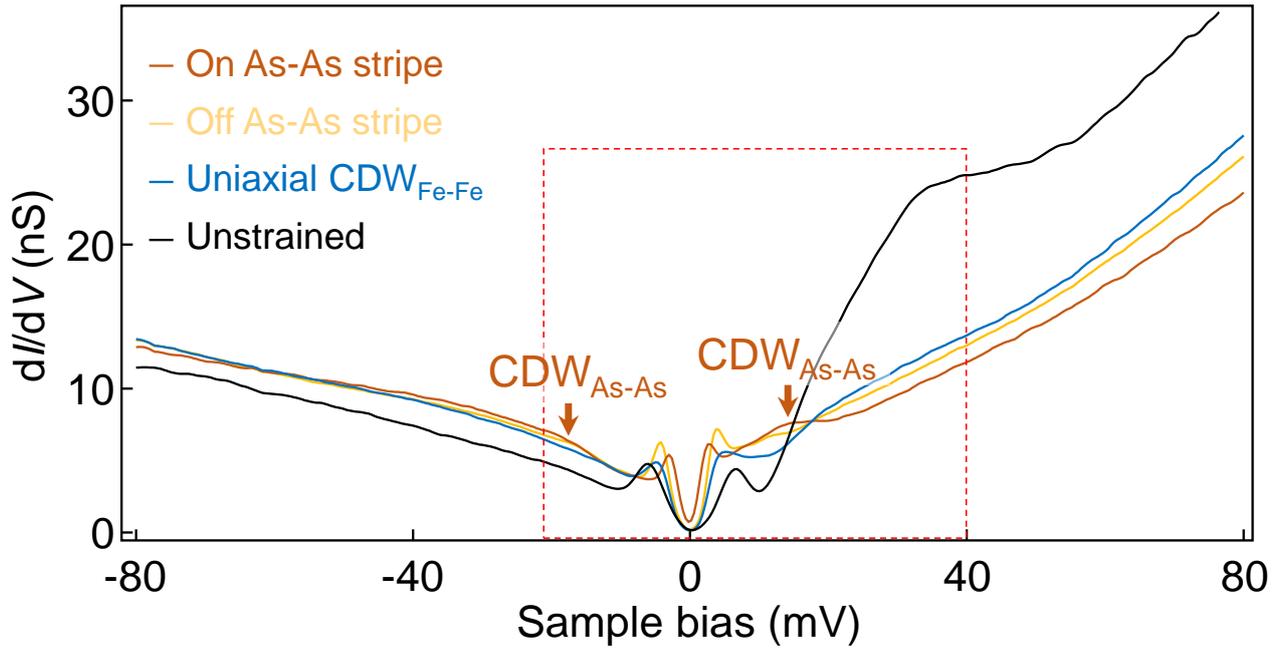

**Extended Data Fig. 14 | d$I$/d$V$ spectra of the on As-As stripe, off As-As stripe, uniaxial CDW$_{Fe\text{-}Fe}$ and unstrained regions over a wide energy range.** The red dashed rectangle outlines the energy window of Fig. 3d.

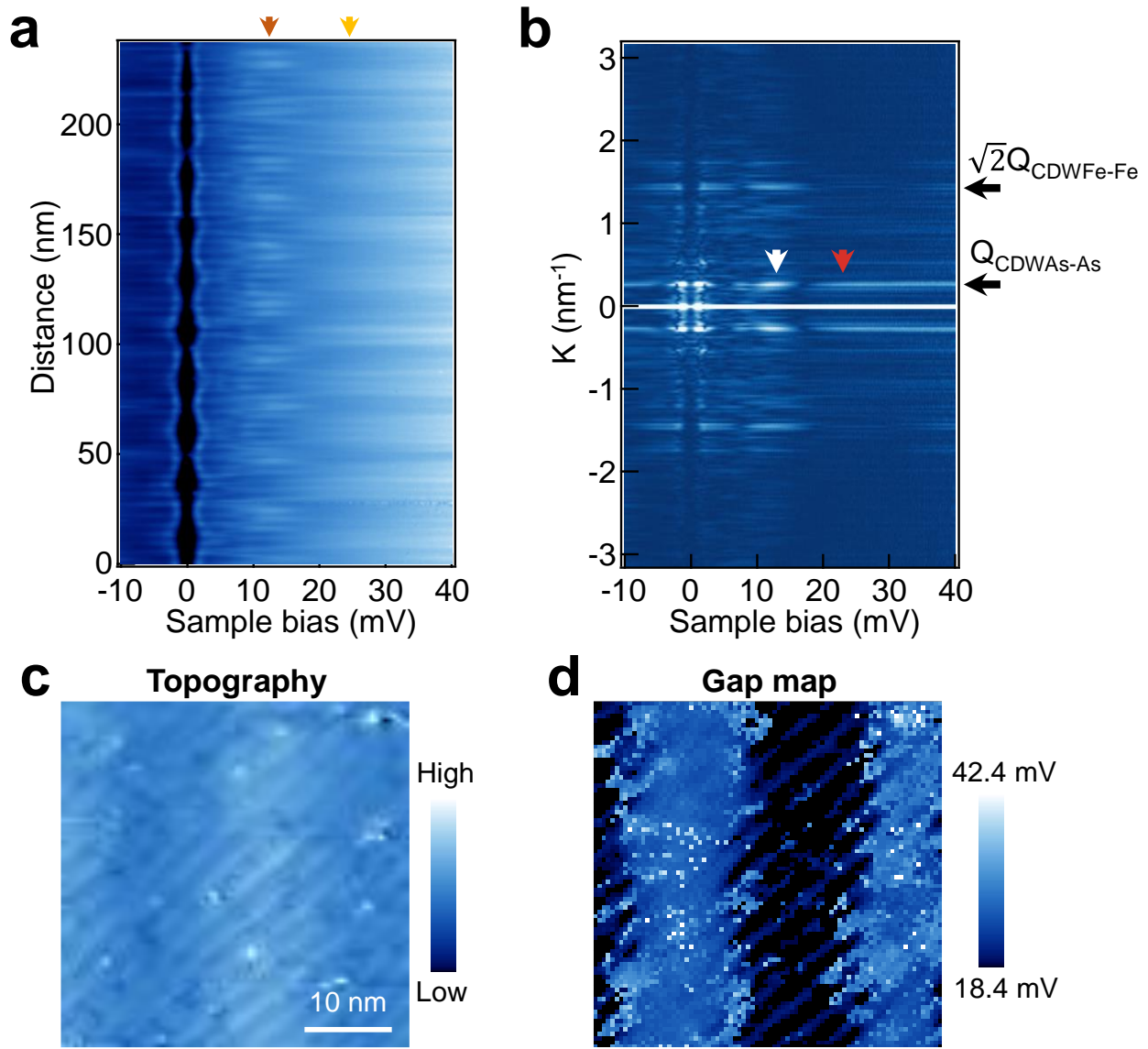

**Extended Data Fig. 15 | Spatially dependent modulation of the CDW gap in the strained region. a,b,** d$I$/d$V$ spectra across the strained region (a) and the corresponding FT image (b). The hump features at energies of ~13 mV and ~22 mV are modulated by the As-As stripes. **c,d,** STM image (c) and CDW$_{As-As}$ gap map (d) of the biaxial CDW region. The CDW gap is extracted by calculating the peak-to-peak values of the CDW coherence peaks as labeled in Extended Data Fig. 14. The gap value is strongly modulated by the As-As stripes in a way that the gap sizes on the As-As stripes are lower than off the stripes (in c: $V_s$=-15 mV, $I_t$=200 pA).

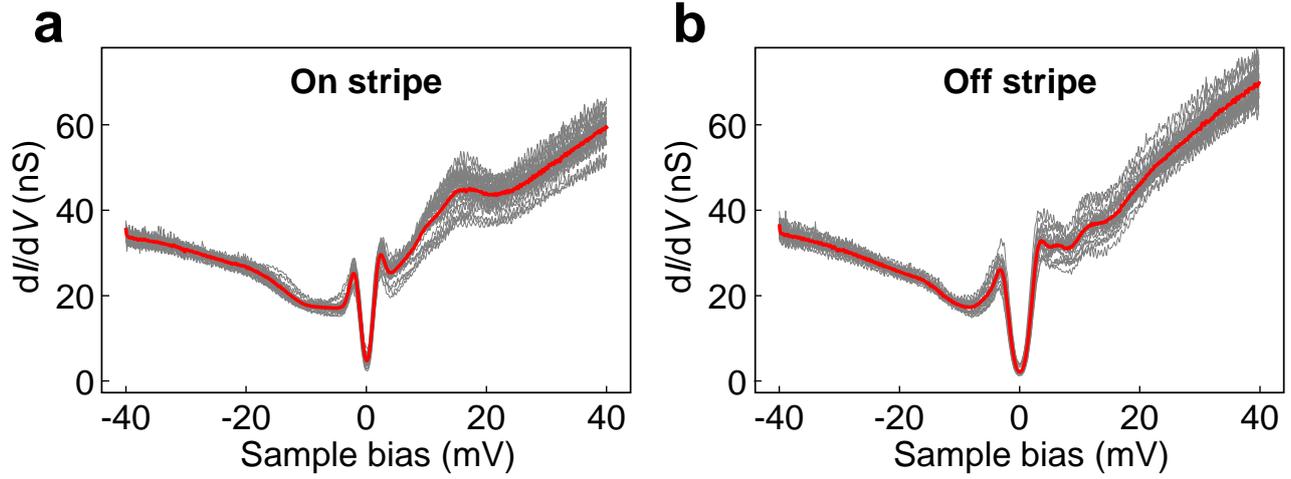

**Extended Data Fig. 16 | d$I$/d$V$ spectra on (a) and off (b) bright As-As stripes, showing V-shaped profiles with minima at ~-10 mV.**

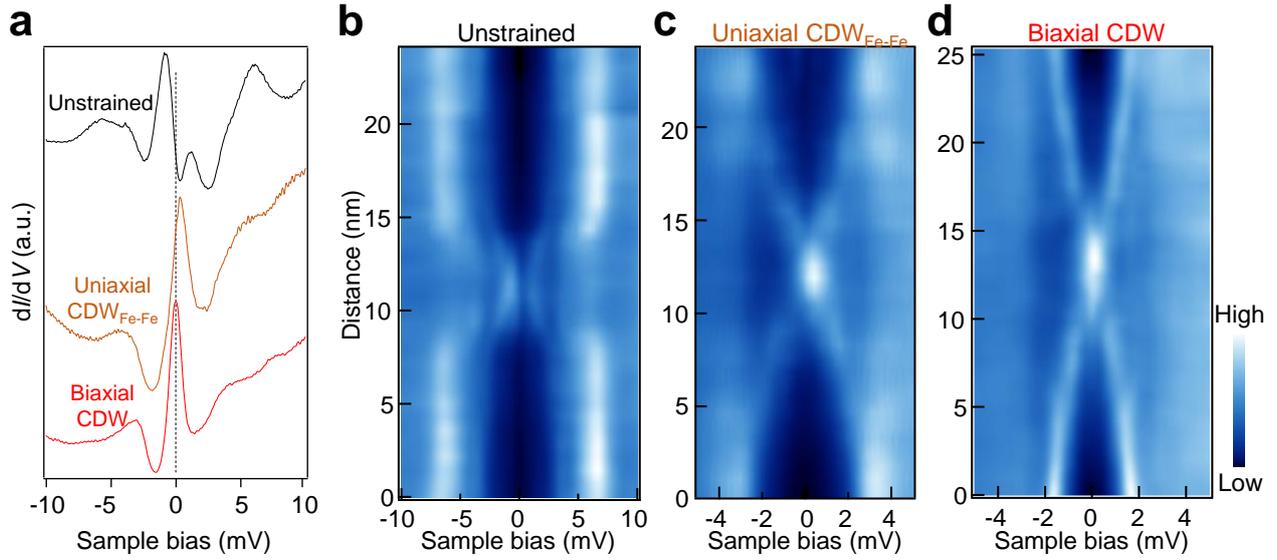

**Extended Data Fig. 17 | Comparison of the spectral feature of the vortices in the unstrained, uniaxial CDW$_{Fe-Fe}$ and biaxial CDW regions. a,** d$I$/d$V$ spectra taken at the centers of ordinary vortices in the unstrained (black), the uniaxial CDW$_{Fe-Fe}$ region (brown) and a topological vortex in the biaxial CDW region (red). **b,** Intensity map of the d$I$/d$V$ linecut across the ordinary vortex in the unstrained region. **c,** Intensity map of the d$I$/d$V$ linecut across the ordinary vortex in the uniaxial CDW$_{Fe-Fe}$ region. **d,** Intensity map of the d$I$/d$V$ linecut across the topological vortex in the biaxial CDW region. The topological vortices exist only in the biaxial CDW regions.

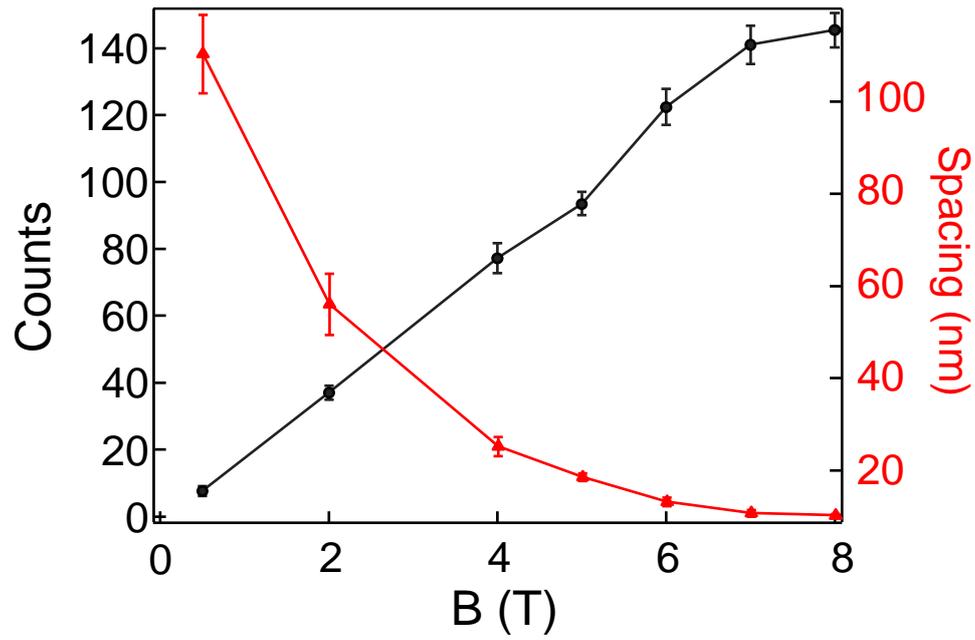

**Extended Data Fig. 18 | Number (black) and the spacing of neighboring vortices (red) under different magnetic fields.** The scanning area is 240 nm × 240 nm.

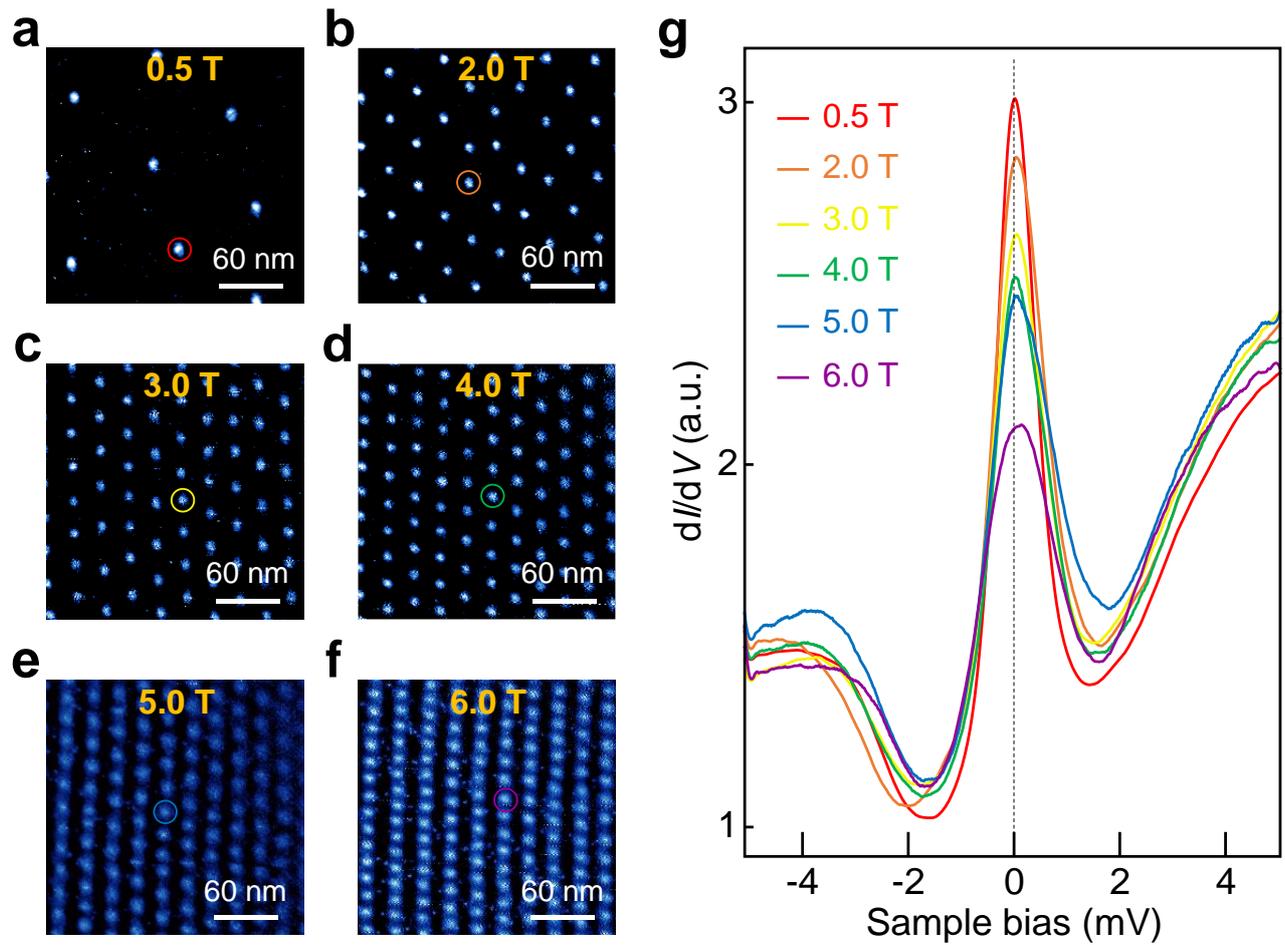

**Extended Data Fig. 19 | Correlation between the vortex spacing and the d$I$/d$V$ spectra of the MZM vortices. a-f,** d$I$/d$V$ maps of MZM lattices at 0 mV under magnetic fields from 0.5 T to 6 T, respectively. **g,** Averaged d$I$/d$V$ spectra under different magnetic fields. The spectra are taken under the same scanning settings. With increasing magnetic fields, the ZBCPs of the d$I$/d$V$ spectra get lower and broader. This phenomenon indicates that a coupling of the MZMs appears when the vortices get closer to each other under higher magnetic fields. The averaged d$I$/d$V$ spectra under different fields can be found in Supplementary Video 3.